\documentclass[a4paper,10pt]{article}

\usepackage{amsthm,amsmath,amssymb,amsfonts}

\def \ring {{\cal R}}
\def \rp {{\cal R}_+}

\def\bbbc{{\mathbb C}}
\def\bbbz{{\mathbb Z}}

\def\hu{{\hat{u}}}

\def\s3{\sqrt{3}}
\def\s5{\sqrt{5}}

\newtheorem{Def}{Definition}
\newtheorem{The}{Theorem}
\newtheorem{Pro}{Proposition}

\begin{document}
\title{Generalisations of the Camassa-Holm equation}
\author{Vladimir Novikov}
\maketitle

\begin{center}
Department of Mathematical Sciences \\ Loughborough University \\
Loughborough, Leicestershire LE11 3TU \\ United Kingdom \\[2ex]
e-mail: \\[1ex]
\texttt{V.Novikov@lboro.ac.uk}
\end{center}

\abstract{We classify  generalised Camassa-Holm type equations which
possess infinite hierarchies of higher symmetries. We show that the
obtained equations can be treated as negative flows of integrable
quasi-linear scalar evolution equations of orders 2, 3 and 5. We
present the corresponding Lax representations or linearisation
transformations for these equations. Some of the obtained equations
seem to be new.}
\section{Introduction}

In recent years there has been a growing interest in integrable
non-evolutionary partial differential equations of  the form
\begin{equation}
\label{CHt} (1-D_x^2)u_t=F(u,u_x,u_{xx},u_{xxx},\ldots), \quad
u=u(x,t),\quad D_x=\frac{\partial}{\partial x}.
\end{equation}
Here $F$ is some function of $u$ and its derivatives with respect to
$x$. The most celebrated example of this type of equations is the
Camassa--Holm equation \cite{CH}:
\[
(1-D_x^2)u_t=3uu_x-2u_xu_{xx}-uu_{xxx}.
\]
Another equivalent form of the Camassa--Holm equation is
\[
m_t=2mu_x+um_x,\quad m=u-u_{xx}.
\]
The Camassa--Holm equation is integrable by the inverse scattering
transform. It possesses an infinite hierarchy of local  conservation
laws, bi-Hamiltonian structure and other remarkable properties of
integrable equations. Despite its non-evolutionary form the
Camassa--Holm equation possesses an infinite hierarchy of local
higher symmetries - indeed this equation can be viewed as an inverse
flow of the equation $u_{\tau}=D_x (u-u_{xx})^{-\frac{1}{2}}$.
Furthermore, the Camassa--Holm equation can be reduced via a
reciprocal transformation to the first negative of the Korteweg--de
Vries hierarchy (see also \cite{Fuchss}). The Camassa--Holm equation possesses multi-phase peakon solutions (peaked soliton solutions with discontinuous derivatives at the peaks).

Until 2002 the Camassa-Holm equation was the only known
integrable
example of the type (\ref{CHt}), which possesses peakon solutions, when Degasperis and Procesi isolated
another equation
\[
(1-D_x^2)u_t=4uu_x-3u_xu_{xx}-uu_{xxx},
\]
or in a different form
\[
m_t=3mu_x+um_x,\quad m=u-u_{xx},
\]
which was also found to be integrable by the inverse scattering
transform \cite{DHH}. The Degasperis--Procesi equation also
possesses infinitely many conservation laws, bi-Hamiltonian
structure etc. It also possesses an infinite hierarchy of local
higher symmetries and can be seen as a non-local symmetry of a local
evolutionary equation $u_{\tau}=(4-D_x^2)D_x
(u-u_{xx})^{-\frac{2}{3}}$. In fact, the Degasperis--Procesi
equation can be reduced via a reciprocal transformation to the first
negative flow of the Kaup-Kupershmidt hierarchy \cite{HW1}.

One may ask questions: are there other integrable equations of the
form (\ref{CHt}) and is it possible to classify all integrable
equations of this type? The answer to both questions is positive.

The first classification result of equations of type (\ref{CHt}) was
obtained in \cite{MN} using the perturbative symmetry approach in
the symbolic representation. In the symmetry approach the existence
of infinite hierarchies of higher symmetries is adopted as a
definition of integrability. The conditions of existence of higher
symmetries are very restrictive and result in algorithmic and
efficient integrability test. In particular, the following result
was proved in \cite{MN}:
\begin{The}
If equation
\[
m_t=b mu_x+um_x,\quad m=u-u_{xx},\quad b\in\bbbc\setminus\{0\}
\]
possesses an infinite hierarchy of (quasi-) local higher symmetries
then $b=2,3$.
\end{The}
Obviously, the case $b=2$ corresponds to the Camassa--Holm equation,
while $b=3$ gives the Degasperis--Procesi equation.

In this paper we extend the classification result of \cite{MN} and
apply the perturbative symmetry approach to isolate and classify
more general class of integrable equations of the form (\ref{CHt}).
We assume that function $F$ in the right hand side is a homogeneous
differential polynomial over $\bbbc$, quadratic or cubic in $u$ and
its $x$-derivatives. The obtained list  comprises of 28 equations
(see section 3) and some of these equations seem to be new to the
best of our knowledge.  The list includes an equation of the form
\[
(1-D_x^2)u_t=u^2u_{xxx}+3uu_xu_{xx}-4u^2u_x.
\]
Integrability and multipeakon solutions of this equation have been
recently studied in \cite{HW} and \cite{HLS}.  For all the obtained
equations we present their first non-trivial higher symmetries. We
also give Lax representations or linearisation transformations for
most of the equations. We show that all the obtained equations can
be treated as negative flows of integrable quasi-linear scalar
evolution equations of orders 2, 3 or 5. The  classification results
of the latter ones can be found in \cite{MSS}.

\section{Integrability test}

In this section we briefly remind the basic definitions and notations of the perturbative symmetry approach (for details see \cite{MN}, \cite{MNWr}). We also present the integrability test  \cite{MN}, which we apply to isolate integrable generalisations of  the Camassa--Holm equation.

\subsection{Symmetries and approximate symmetries}

In what follows we shall consider the Camassa--Holm type equation
(\ref{CHt}) with the right hand side being a differential polynomial
over $\bbbc$.

Let $\ring$ be a ring of differential polynomials in
$u,u_x,u_{xx},\ldots$ over $\bbbc$. We shall adopt a notation
\[
u_i\equiv D_x^{i}(u).
\]
We shall often omit subscript $0$ at  $u_0$ and write $u$ instead of $u_0$.

The ring $\ring$ is a differential ring with a derivation
\[
D_x=\sum_{i\ge 0}u_{i+1}\frac{\partial}{\partial u_i}.
\]
The ring has a natural gradation with respect to degrees of
non-linearity in $u$ and its $x$-derivatives:
\[
\ring=\bigoplus_{i\ge 0}\ring_i,\quad
\ring_i=\{f(u,u_1,\ldots,u_k)\in\ring\,|\,f(\lambda u,\lambda
u_1,\ldots,\lambda u_k)=\lambda^if(u,u_1,\ldots,u_k)\},\,\,\,
\lambda\in\bbbc.
\]
The space $\ring_0=\bbbc$, $\ring_1$ is a space of linear
polynomials in $u,u_1,\ldots$, $\ring_2$ is a space of quadratic
polynomials etc. It is convenient to introduce a notion of ``little
oh" as
\[
f=o(\ring_p)\Leftrightarrow f\in\bigoplus_{i>p}\ring_i.
\]

Let us denote by $\rp$  a differential ring without a unit
\[
\rp=\bigoplus_{i> 0}\ring_i.
\]

Suppose that the right hand side  of the equation (\ref{CHt})
$F\in\rp$. We can formally rewrite equation (\ref{CHt}) as an
evolutionary equation as
\begin{equation}
\label{CHg} u_t=\Delta(F),\quad \Delta=(1-D_x^2)^{-1}.
\end{equation}
Symmetries and conservation laws of this equation, if they exist,
may also contain operator $\Delta$ in their structure and therefore
we need an extension of the differential ring $\rp$ with the
operator $\Delta$. The construction of such extension was first
suggested in \cite{MY} for the evolutionary $2+1$ dimensional
equations. For the Camassa--Holm type equations it was first applied
in \cite{MN}. Namely, let us construct a sequence of spaces
$\rp^i,\,\,i=0,1,2,\ldots$ as follows:
\[
\rp^0=\rp,\quad \rp^1=\overline{\rp^0\bigcup \Delta(\rp^0)},\quad
\rp^{n+1}=\overline{\rp^n\bigcup \Delta(\rp^n)}.
\]
The subscript $n$ in $\rp^n$ is the ``nesting depth" of the operator
$\Delta$. The extension construction is compatible with the natural
gradation:
\[
\rp^n=\bigoplus_{i>0}\ring_i^n,\quad
\ring_i^n=\{f[u]\in\ring_+^n\,|\,f[\lambda u]=\lambda^if[u]\},\,\,\,
\lambda\in\bbbc.
\]

It is clear that $\Delta(F)$ in the equation (\ref{CHg}) belongs to
$\rp^1$. The symmetries of the equation may belong to $\rp^k$ for
some appropriate $k\ge 0$ and we introduce the following definition
of a symmetry:

\begin{Def} A function $G\in\rp^k,\,\, k\ge 0$ is called a generator of a symmetry of equation
(\ref{CHg}) if a differential equation
\[
u_{\tau}=G
\]
is compatible with the equation (\ref{CHg}): $G_t-F_{\tau}=0$.
\end{Def}

We adopt the following definition of integrability:
\begin{Def} Equation (\ref{CHg}) is integrable if it possesses an
infinite hierarchy of symmetries.
\end{Def}

In addition to the definition of a symmetry we also introduce a
definition of an approximate symmetry:

\begin{Def}
 A function $G\in\rp^k,\,\, k\ge 0$ is called a generator of an approximate symmetry of degree $p$ of equation
(\ref{CHg}) if $G_t-F_{\tau}=o(\ring_p^k)$.
\end{Def}

Any equation
\[
u_t=\Delta(F)=\Delta(F_1)+\Delta(F_2)+\cdots+\Delta(F_k),\quad
F_k\in\ring_k
\]
possesses an infinite hierarchy of approximate symmetries of degree
1 -- these are symmetries of its linear part $u_t=\Delta(F_1)$. The
condition of existence of approximate symmetries of degree 2 imposes
strong restrictions on the equation. However an equation may possess
infinitely many of approximate symmetries of degree 2, but fail to
possess approximate symmetries of degree 3. On the other hand an
integrable equation possesses infinitely many approximate symmetries
of any degree. The degree of approximate symmetry can be viewed as a
measure to the integrability. In many cases the existence of
approximate symmetries of sufficiently large degree implies
integrability.

In order to derive the conditions of existence of symmetries and
approximate symmetries it is convenient to introduce the symbolic
representation of the ring $\rp$ and its extension.

\subsection{Symbolic representation}

We start by introducing the symbolic representation $\hat{\ring}_+$
of $\rp$. We first introduce the symbolic representation of spaces
$\ring_{k},\,\, k=1,2,\ldots$:

\begin{itemize}
\item[1)] To a linear monomial $u_i\in\ring_{1}$ we put into correspondence a
symbol $$u_i\longrightarrow\hu\,\xi_1^i.$$
\item[2)] To a quadratic monomial $u_iu_j\in\ring_{2}$ we put into
correspondence a symbol
$$u_iu_j\longrightarrow\frac{\hu^2}{2}(\xi_1^i\xi_2^j+\xi_1^j\xi_2^i).$$
\item[3)] We represent a generic $u_0^{n_0}u_1^{n_1}\cdots
u_k^{n_k}\in\ring_n,\,\,n=n_0+n_1+\cdots+n_k$ by a symbol
\[
u_0^{n_0}u_1^{n_1}\cdots
u_k^{n_k}\longrightarrow\hu^n\langle\xi_1^0\cdots\xi_{n_0}^0\xi_{n_0+1}^1\cdots\xi_{n_0+n_1}^1\cdots\xi_n^k\rangle,
\]
where brackets $\langle*\rangle$ denote a symmetrisation operation:
\[
\langle f(\xi_1,\ldots,\xi_n)\rangle=\sum_{\sigma\in
S_n}f(\xi_{\sigma(1)},\ldots,\xi_{\sigma(n)}).
\]
\end{itemize}
We define addition, multiplication and derivation as follows. Let
$f\in\ring_i,\,g\in\ring_j$ be two monomials and their symbolic
representation is given by $f\to\hu^i a(\xi_1,\ldots,\xi_i)$,
$g\to\hu^j b(\xi_1,\ldots,\xi_j)$. Then
\[
f+g\longrightarrow \hu^i a(\xi_1,\ldots,\xi_i)+\hu^j
b(\xi_1,\ldots,\xi_j)
\]
and
\[
f\cdot g\longrightarrow \hu^{i+j}\langle a(\xi_1,\ldots,\xi_i)
b(\xi_{i+1},\ldots,\xi_{i+j})\rangle.
\]
In particular, if $i=j$ then
$f+g\to\hu^i\left(a(\xi_1,\ldots,\xi_i)+b(\xi_1,\ldots,\xi_i)\right)$.

To a derivative of $f\to\hu^i a(\xi_1,\ldots,\xi_i)$  we put into correspondence
\[
D_x(f)\longrightarrow \hu^i
a(\xi_1,\ldots,\xi_i)(\xi_1+\cdots+\xi_i).
\]
This concludes the construction of the symbolic representation
$\hat{\ring}_+$ of the differential ring $\rp$.

We also introduce a notion of a pseudo-differential formal series
in the symbolic representation. We reserve a special symbol $\eta$
for the operator $D_x$ in the symbolic representation with an action
rule
\[
\eta(\hu^n a(\xi_1,\ldots,\xi_n))=\hu^n
a(\xi_1,\ldots,\xi_n)(\xi_1+\cdots+\xi_n).
\]
Let $fD_x^p$ and $gD_x^q$, $p,q\in\bbbz$ be two
(pseudo)-differential operators and suppose that $f\to
\hu^ia(\xi_1,\ldots,\xi_i)$, $g\to \hu^jb(\xi_1,\ldots,\xi_j)$. Then
for the symbolic representation of these operators we have
\[
fD_x^p\longrightarrow \hu^ia(\xi_1,\ldots,\xi_i)\eta^p,\quad
gD_x^q\longrightarrow \hu^jb(\xi_1,\ldots,\xi_i)\eta^q.
\]
For the addition and composition of pseudo-differential operators in
the symbolic representation we have
\[
fD_x^p+gD_x^q\longrightarrow
\hu^ia(\xi_1,\ldots,\xi_i)\eta^p+\hu^jb(\xi_1,\ldots,\xi_i)\eta^q
\]
\[
fD_x^p\circ gD_x^q\longrightarrow \hu^{i+j}\langle
a(\xi_1,\ldots,\xi_i)(\eta+\xi_{i+1}+\cdots+\xi_{i+j})^p
b(\xi_{i+1},\ldots,\xi_{i+j})\eta^q\rangle.
\]

More generally we shall consider formal series in the form
\begin{equation}
\label{ser} A=a_0(\eta)+\hu
a_1(\xi_1,\eta)+\hu^2a_2(\xi_1,\xi_2,\eta)+\hu^3a_3(\xi_1,\xi_2,\xi_3,\eta)+\cdots
\end{equation}
where functions $a_k(\xi_1,\ldots,\xi_k,\eta)$ are symmetric
functions with respect to arguments $\xi_1,\ldots,\xi_k$. The
addition rule of such series is obvious while for  composition of
two monomials we have
\[
\hu^ia(\xi_1,\ldots,\xi_i,\eta)\circ
\hu^jb(\xi_1,\ldots,\xi_j,\eta)=\hu^{i+j}\langle
a(\xi_1,\ldots,\xi_i,\eta+\xi_{i+1}+\cdots+\xi_{i+j})b(\xi_{i+1},\ldots,\xi_{i+j},\eta)\rangle
\]
where the symmetrisation operation is taken with respect to all
arguments $\xi_1,\ldots,\xi_{i+j}$, but not $\eta$.

We introduce a notion of {\it locality} of a pseudo-differential
operator
\begin{Def}
Function $a(\xi_1,\ldots,\xi_i,\eta)$ is called local if all
coefficients $a_j(\xi_1,\ldots,\xi_i)$ of its expansion in $\eta$ at
$\eta\to\infty$
\[
a(\xi_1,\ldots,\xi_i,\eta)=\sum_{j<s}a_j(\xi_1,\ldots,\xi_i)\eta^j
\]
are symmetric polynomials in variables $\xi_1,\ldots,\xi_i$. Formal
series (\ref{ser}) is called local if all functions
$a_j(\xi_1,\ldots,\xi_j)$, $j=1,2,\ldots$ in (\ref{ser}) are local.
\end{Def}

To construct the symbolic representation of the extension of the
ring $\rp$ with the operator $\Delta=(1-D_x^2)^{-1}$ it is enough to
note that the symbolic representation of the operator $\Delta$ is
\[
\Delta\longrightarrow (1-\eta^2)^{-1}.
\]
Indeed, if $f\in\ring_k$ and $f\to \hu^ka(\xi_1,\ldots,\xi_k)$, then
\[
\Delta(f)\longrightarrow
\hu^k\frac{a(\xi_1,\ldots,\xi_k)}{1-(\xi_1+\cdots+\xi_k)^2} .
\]
Using if necessary the addition and multiplication operations  we
thus can obtain the symbolic representation of any space
$\ring_+^j$.

In addition to the notion of locality of a pseudo-differential series we also introduce a notion of {\it quasi-locality}

\begin{Def} A pseudo-differential operator
$$
\hu^na(\xi_1,\ldots,\xi_n,\eta)=\sum_{i<s}\hu^na_i(\xi_1,\ldots,\xi_n)\eta^i
$$
is called quasi-local if for all $i<s$ $\hu^na_i(\xi_1,\ldots,\xi_n)$ are symbolic representations of some elements from $\ring_n^k$ for some $k\ge 0$. A formal series (\ref{ser}) is called quasi-local if all its terms are quasi-local.
\end{Def}

Finally we introduce a notion of a Frechet derivative in the
symbolic representation: let $f\in\ring_k^n,\,\,k>0,n\ge 0$ and its
symbolic representation is given by
$f\to\hat{f}=\hu^ka(\xi_1,\ldots,\xi_k)$. Then to the Frechet
derivative $f_*$ corresponds:
\[
f_*\to\hat{f}_*=k\hu^{k-1}a(\xi_1,\ldots,\xi_{k-1},\eta).
\]

\subsection{Symmetries and approximate symmetries in the symbolic
representation}

Now we derive conditions of existence of symmetries and approximate
symmetries of equation (\ref{CHg}). We shall suppose that $F\in\rp$
and thus we can rewrite equation (\ref{CHg}) as
\begin{equation}
\label{CHex}
u_t=\Delta(F)=\Delta(F_1)+\Delta(F_2)+\cdots+\Delta(F_k),\quad
F_i\in\ring_i,\,\,i=1,2,\ldots .
\end{equation}
We write the symbolic representation of $\Delta(F)$ as
\begin{equation}
\label{Fs} \Delta(F)\longrightarrow \hat{F}= \hu \omega(\xi_1)+\hu^2
a_1(\xi_1,\xi_2)+\cdots+\hu^k a_{k-1}(\xi_1,\ldots,\xi_k).
\end{equation}
By construction $a_i(\xi_1,\ldots,\xi_{i+1}),\,\,i=1,\ldots,k-1$ are
symmetric rational functions in $\xi_1,\ldots,\xi_{i+1}$ of the form
\[
a_i(\xi_1,\ldots,\xi_{i+1})=\frac{b_i(\xi_1,\ldots,\xi_{i+1})}{1-(\xi_1+\cdots+\xi_{i+1})^2}
\]
where symmetric polynomials $b_i(\xi_1,\ldots,\xi_{i+1})$ are
symbolic representations of differential polynomials $F_{i+1}$,
$i=1,2,\ldots,k-1$. Similarly
$\omega(\xi_1)=\tilde{\omega}(\xi_1)/(1-\xi_1^2)$ and
$\tilde{\omega}(\xi_1)$ a symbolic representation of $F_1$. We shall
suppose that $F_1$ is such that $\omega(\xi_1)\ne const\,\xi_1$.

Let $G\in\ring_+^n,n\ge 0$ is a symmetry of (\ref{CHex}). Without
loss of generality we can suppose that
\[
G=G_1+G_2+\ldots+G_m,\quad G_i\in\ring^n_i,\,\,i=1,\ldots,m.
\]
Let
\begin{equation}
\label{sym} G\longrightarrow \hu \Omega(\xi_1)+\hu^2
A_1(\xi_1,\xi_2)+\cdots+\hu^k A_{m-1}(\xi_1,\ldots,\xi_m)
\end{equation}
be a symbolic representation of $G$, i.e.
$\hu^{i}A_{i-1}(\xi_1,\ldots,\xi_{i}),\,\,i=1,\ldots,m-1$ are
symbolic representations of $G_i\in\ring^n_i$ and thus are symmetric
rational functions in $\xi_1,\ldots,\xi_i$.

The following proposition holds:
\begin{Pro} Function $G\in\ring_+^n,n\ge 0$ with the symbolic representation (\ref{sym}) is
a generator of a symmetry of equation (\ref{CHex}) with the symbolic
representation (\ref{Fs}) if and only if
\begin{eqnarray}
\nonumber
A_1(\xi_1,\xi_2)&=&\frac{\Omega(\xi_1+\xi_2)-\Omega(\xi_1)-\Omega(\xi_2)}
{\omega(\xi_1+\xi_2)-\omega(\xi_1)-\omega(\xi_2)}a_1(\xi_1,\xi_2),\\
\nonumber A_m(\xi_1,...,\xi_{m+1})&=&\frac{G^\Omega
(\xi_1,...,\xi_{m+1})}{G^\omega(\xi_1,...,\xi_{m+1})}a_m(\xi_1,...,\xi_{m+1})+\\
\nonumber &&G^\omega (\xi_1,...,\xi_{m+1})^{-1}\cdot\bigg[
\langle\sum_{j=1}^{m-1}\frac{m+1}{m-j+1}A_j(\xi_1,...,\xi_j,\sum_{k=j+1}^{m+1}
\xi_k)a_{m-j}(\xi_{j+1},...,\xi_{m+1})-
\\
\nonumber &&
 -\sum_{j=1}^{m-1}\frac{m+1}{j+1} a_{m-j} (\xi_1,...,
\xi_{m-j},\sum_{k=m-j+1}^{m+1}\xi_k)\cdot
A_j(\xi_{m-j+1},...,\xi_{m+1}) \rangle\bigg]
\end{eqnarray}
where
\[
G^\omega (\xi_1,...,\xi_m)=\omega(\sum_{n=1}^{m}\xi_n)-
\sum_{n=1}^{m}\omega(\xi_n),\quad\quad G^\Omega
(\xi_1,...,\xi_m)=\Omega(\sum_{n=1}^{m}\xi_n)-\sum_{n=1}^{m}\Omega(\xi_n)\,
\]
and $\hu^iA_{i-1}(\xi_1,\ldots,\xi_{i-1})$ are symbolic
representations of elements of $\ring_i^n$.
\end{Pro}
The proof follows from the compatibility conditions of equations
(\ref{CHex}) and $u_{\tau}=G$ (for details see \cite{MN}).
Proposition 1 gives necessary and sufficient conditions of existence
of an approximate symmetry of degree $p$. Indeed, if for a given
equation (\ref{CHex}) with the symbolic representation (\ref{Fs})
$\hu^iA_{i-1}(\xi_1,\ldots,\xi_i)$ are symbolic representations of
elements of $\ring_i^n$ for all $i=1,2,\ldots,p$ then $G$ is an
approximate symmetry of degree $p$. Note that if $G$ is a symmetry
then it is completely determined by its linear part $G_1$. From
proposition 1 it follows that to characterise a hierarchy of
symmetries it is sufficient to charactarise a hierarchy of
admissable linear terms.

However it is possible to derive  the necessary conditions of
existence of an infinite hierarchy of (approximate) symmetries
without knowing the structure of admissible linear terms of the
symmetries. To do so we introduce a notion of a formal recursion
operator:

\begin{Def} A quasi-local formal series
\begin{equation}
\label{L}
\Lambda=\phi(\eta)+\hu\phi_1(\xi_1,\eta)+\hu^2\phi_2(\xi_1,\xi_2,\eta)+\hu^3\phi_3 (\xi_1,\xi_2,\xi_3,\eta )+\cdots
\end{equation}
is called a formal recursion operator for equation (\ref{CHex}) if
it satisfies
\begin{equation}
\label{Leq}
\Lambda_t= \hat{F}_*\circ\Lambda-\Lambda\circ\hat{F}_*
\end{equation}
where $\hat{F}_*$ is a symbolic representation of a Frechet derivative of $F$.
\end{Def}

The following statement holds:
\begin{The}\label{TheL}
If equation (\ref{CHex}) possesses an infinite hierarchy of higher symmetries then it possesses a formal recursion operator (\ref{L}) with $\phi(\eta)=\eta$.
\end{The}
The proof of the theorem can be found in \cite{MN}.

Equation $\Lambda_t=\hat{F}_*\circ\Lambda-\Lambda\circ\hat{F}_*$ can be resolved in terms of functions $\phi_i(\xi_1,\ldots,\xi_i,\eta)$:
\begin{Pro} \label{prolambda} Let $\phi(\eta )$ be an arbitrary function and
formal series
\begin{equation*}
\label{Lsym} \Lambda=\phi(\eta )+\hu\phi_1 (\xi_1,\eta )+\hu^2\phi_2
(\xi_1,\xi_2,\eta )+\hu^3\phi_3 (\xi_1,\xi_2,\xi_3,\eta )+\cdots
\end{equation*}
be a solution of equation (\ref{Leq}), then its coefficients
$\phi_m(\xi_1,...,\xi_m,\eta )$ can be found recursively
\begin{eqnarray*}
&&\phi_1 (\xi_1,\eta )=\frac{2(\phi(\eta+\xi_1)-\phi(\eta ))}{G^\omega(\xi_1,\eta )}
a_1(\xi_1,\eta )
\\   &&\phi_m
(\xi_1,...,\xi_m,\eta )=\frac{1}{G^\omega (\xi_1,...,\xi_m,\eta )} \bigg((m+1)
(\phi(\eta+\xi_1+... +\xi_m) -\phi(\eta ))a_m(\xi_1,...,\xi_m,\eta )\\&&+\sum_{n=1}^{m-1}\langle n
\phi_n(\xi_1,..,\xi_{n-1},\xi_n+\cdots +\xi_m,\eta ) a_{m-n}(\xi_n,..,\xi_m)+(m-n+1) \phi_n(\xi_1,..,\xi_{n},\eta+\sum_{l=n+1}^m\xi_{l})
a_{m-n}(\xi_{n+1},..,\xi_m,\eta )\\  &&
-(m-n+1)a_{m-n}(\xi_{n+1},..,\xi_m,\eta+\sum_{l=1}^n\xi_{l})
\phi_n(\xi_{1},..,\xi_n,\eta ) \rangle \bigg).
\end{eqnarray*}
\end{Pro}
The proof can be found in \cite{MN}.

Theorem \ref{TheL} and proposition \ref{prolambda} suggest the following integrability test for equation (\ref{CHex}):
\begin{itemize}
\item Compute the symbolic representation of equation (\ref{CHex}) and calculate the first few coefficients $\phi_i(\xi_1,\ldots,\xi_i,\eta),\,\, i=1,2,\ldots$;
\item Check the quasi-locality conditions
\end{itemize}

In the next section we apply this test to isolate and classify integrable generalisations of the Camassa--Holm equation.

\section{Lists of generalised Camassa--Holm type equations}

In this section we present the classification results of Camassa--Holm type equations  with quadratic and cubic non-linearity. We consider the following three ans\"{a}tze for equation (\ref{CHex}):
\begin{eqnarray}
\nonumber
(1-\epsilon^2D_x^2)u_t&=&c_1uu_x+\epsilon\left[c_2uu_{xx}+c_3u_x^2\right]+\epsilon^2\left[c_4uu_{xxx}+c_5u_xu_{xx}\right]
+\epsilon^3\left[c_6uu_{xxxx}+c_7u_xu_{xxx}+c_8u_{xx}^2\right]\\ \label{e1}
&+&\epsilon^4\left[c_9uu_{xxxxx}+
c_{10}u_xu_{xxxx}+c_{11}u_{xx}u_{xxx}\right],
\end{eqnarray}

\begin{eqnarray}
\label{e2}
(1-\epsilon^2D_x^2)u_t&=&c_1u_x^2+\epsilon c_2u_xu_{xx}+\epsilon^2[c_3u_xu_{xxx}+c_4u_{xx}^2]+
\epsilon^3[c_5u_xu_{xxxx}+c_6u_{xx}u_{xxx}]\\ \nonumber &+&\epsilon^4[c_7u_xu_{xxxxx}+c_8u_{xx}u_{xxxx}+c_9u_{xxx}^2]
\end{eqnarray}
and

\begin{eqnarray}
\label{e3}
(1-\epsilon^2D_x^2)u_t&=&c_1u^2u_x+\epsilon[c_2u^2u_{xx}+c_3uu_x^2]+\epsilon^2[c_4u^2u_{xxx}+c_5uu_xu_{xx}+c_6u_x^3]\\
\nonumber &+&
\epsilon^3[c_7u^2u_{xxxx}+c_8uu_xu_{xxx}+c_9uu_{xx}^2+c_{10}u_x^2u_{xx}]\\
\nonumber
&+&\epsilon^4[c_{11}u^2u_{xxxxx}+c_{12}uu_xu_{xxxx}+c_{13}uu_{xx}u_{xxx}+
c_{14}u_x^2u_{xxx}+c_{15}u_xu_{xx}^2].
\end{eqnarray}
Here $\epsilon$ and $c_i$ are complex parameters and $\epsilon\ne 0$. The right hand sides of equations (\ref{e1}), (\ref{e2}) and (\ref{e3})
 are homogeneous differential polynomials of weights $1$, $2$ and $1$ respectively if we assume that weight of $u_i$ is $i$ and weight of $\epsilon$ equals $-1$.

\subsection{Equations with quadratic nonlinearity}

\begin{The} Suppose that at least one of the following equations is not satisfied:
\begin{equation}
\label{cond1}
c_{2}=0,\quad c_6=0,\quad c_9=0,\quad c_1+c_4=0.
\end{equation}
Then if  equation (\ref{e1}) possesses an infinite hierarchy of quasi-local higher symmetries
then up to re-scaling $x\to\alpha x, \,t\to\beta t,u\to\gamma u$,
$\alpha,\beta,\gamma=const$ it is one of the list:

\begin{eqnarray}
\label{CH} (1-\epsilon^2D_x^2)u_t&=&3uu_x-2\epsilon^2u_xu_{xx}-\epsilon^2uu_{xxx},\\ \nonumber \\
\label{DP} (1-\epsilon^2D_x^2)u_t&=&D_x\left(4-\epsilon^2D_x^2\right)u^2,\\ \nonumber \\
\label{C3} (1-\epsilon^2D_x^2)u_t&=&D_x\left[(4-\epsilon^2D_x^2)u\right]^2,\\ \nonumber\\
\label{C4} (1-\epsilon^2D_x^2)u_t&=&D_x(2+\epsilon D_x)\left[(2-\epsilon D_x)u\right]^2,\\ \nonumber\\
\label{F2} (1-\epsilon^2D_x^2)u_t&=&D_x(2-\epsilon D_x)(1+\epsilon D_x)u^2,\\ \nonumber\\
\label{F4} (1-\epsilon^2D_x^2)u_t&=&D_x(2-\epsilon D_x)\left[(1+\epsilon D_x)u\right]^2,\\ \nonumber\\
\label{F1} (1-\epsilon^2D_x^2)u_t&=&D_x\left[(2-\epsilon D_x)(1+\epsilon D_x)u\right]^2,\\ \nonumber\\
\label{F3} (1-\epsilon^2D_x^2)u_t&=&D_x(1+\epsilon D_x)\left[(2-\epsilon D_x)u\right]^2,\\ \nonumber\\
\label{B3} (1-\epsilon^2D_x^2)u_t&=&(1-\epsilon^2D_x^2)(\epsilon
uu_{xx}-\frac{1}{2}\epsilon u_x^2+cuu_x),\quad
c\in\bbbc,\\ \nonumber\\
\label{B1} (1-\epsilon^2D_x^2)u_t&=&(1-\epsilon D_x)\left[\epsilon
S(u)S(u_{xx})-\frac{1}{2}\epsilon (S(u_x))^2-\frac{1}{2}c
S(u)S(u_x)\right],\quad S=1+\epsilon D_x.
\end{eqnarray}
\end{The}
The condition that at least one of the equations in  (\ref{cond1})
is not satisfied insures that $\omega(\xi_1)\ne const\, \xi_1$ in
the corresponding symbolic representations of the equation. To prove
the theorem it is sufficient to check the quasi-locality conditions
of $\hu\phi_1(\xi_1,\eta), \hu^2\phi_2(\xi_1,\xi_2,\eta)$ and
$\hu^3\phi_3(\xi_1,\xi_2,\xi_3,\eta)$ of the formal recursion
operator:
\[
\Lambda=\eta+\hu\phi_1(\xi_1,\eta)+\hu^2\phi_2(\xi_1,\xi_2,\eta)+\hu^3\phi_3(\xi_1,\xi_2,\xi_3,\eta).
\]
We do not present here the explicit formulae for these functions as
they are quite cumbersome. One can easily compute them using
proposition \ref{prolambda}.

\begin{The} Suppose that at least one of the following equations is not satisfied:
\[
c_{2}=0,\quad c_5=0,\quad c_7=0,\quad 2c_1+c_3=0.
\]
Then if  equation (\ref{e2}) possesses an infinite hierarchy of quasi-local higher symmetries
then up to re-scaling $x\to\alpha x, \,t\to\beta t,u\to\gamma u$,
$\alpha,\beta,\gamma=const$ it is one of the list:
\begin{eqnarray}
\label{pCH} (1-\epsilon^2D_x^2)u_t&=&3u_x^2-2\epsilon^2u_xu_{xxx}-\epsilon^2u_{xx}^2,\\ \nonumber \\
\label{pDP} (1-\epsilon^2D_x^2)u_t&=&\left(4-\epsilon^2D_x^2\right)u_x^2,\\ \nonumber \\
\label{pC3} (1-\epsilon^2D_x^2)u_t&=&\left[(4-\epsilon^2D_x^2)u_x\right]^2,\\ \nonumber\\
\label{pC4} (1-\epsilon^2D_x^2)u_t&=&(2+\epsilon D_x)\left[(2-\epsilon D_x)u_x\right]^2,\\ \nonumber\\
\label{pF2} (1-\epsilon^2D_x^2)u_t&=&(2-\epsilon D_x)(1+\epsilon D_x)u_x^2,\\ \nonumber\\
\label{pF4} (1-\epsilon^2D_x^2)u_t&=&(2-\epsilon D_x)\left[(1+\epsilon D_x)u_x\right]^2,\\ \nonumber\\
\label{pF1} (1-\epsilon^2D_x^2)u_t&=&\left[(2-\epsilon D_x)(1+\epsilon D_x)u_x\right]^2,\\ \nonumber\\
\label{pF3} (1-\epsilon^2D_x^2)u_t&=&(1+\epsilon D_x)\left[(2-\epsilon D_x)u_x\right]^2.
\end{eqnarray}
\end{The}
To prove the theorem it is again necessary to check the quasi-locality conditions of the first three terms of the corresponding formal recursion operator.

Let us consider now some properties of equations (\ref{CH})-(\ref{B1}) and (\ref{pCH})-(\ref{pF3}).

\underline{Camassa-Holm equation (\ref{CH}).} The equation (\ref{CH}) is the Camassa-Holm equation. It can be rewritten as
\[
m_t=2mu_x+um_x,\quad m=u-\epsilon^2u_{xx}.
\]
The Camassa-Holm equation possesses an infinite hierarchy of {\it local} higher symmetries and the
 first non-trivial local symmetry is
\[
u_{\tau}=D_x (u-\epsilon^2u_{xx})^{-\frac{1}{2}}.
\]

The Lax representation and the bi-Hamiltonian structure can be found in \cite{CH,FF}.

\underline{Degasperi--Procesi equation (\ref{DP}).} The equation (\ref{DP}) is the Degasperis-Procesi equation and it can be rewritten as
\[
m_t=6mu_x+2um_x,\quad m=(1-\epsilon^2D_x^2)u.
\]
The Degasperis-Procesi equation also possesses an infinite hierarchy of local higher symmetries and the first non-trivial such a symmetry is
\[
u_{\tau}=(4-\epsilon^2D_x^2)D_x (u-\epsilon^2u_{xx})^{-\frac{2}{3}}.
\]

The bi-Hamiltonian structure and the Lax representation for the Degasperis-Procesi equation can be found in \cite{DHH}.

\underline{Equation (\ref{C3}).} The first non-trivial symmetry of equation (\ref{C3}) is
\[
u_{\tau}=D_x\left[\left(4-\epsilon^2D_x^2\right)(1-\epsilon^2D_x^2)u\right]^{-\frac{2}{3}}.
\]
Equation (\ref{C3}) can be rewritten as
\[
m_t=D_x\left(m+3u\right)^2,\quad m=u-\epsilon^2u_{xx}.
\]
It is easy to see that the Degasperis-Procesi equation transforms into the equation (\ref{C3}) under the transformation
\[
 u\to(4-\epsilon^2D_x^2)u.
 \]

 The Lax representation for the equation (\ref{C3}) is
 \begin{eqnarray*}
 &&\psi_x-\psi_{xxx}-\lambda\left(4m-\epsilon^2m_{xx}\right)\psi=0,\\
 &&\psi_t=\frac{2}{\lambda}\psi_{xx}+2(m+3u)\psi_x-2(m_x+3u_x+\frac{2}{3\lambda})\psi.
 \end{eqnarray*}

 \underline{Equation (\ref{C4}).} The first non-trivial symmetry of equation (\ref{C4}) is
 \[
 u_{\tau}=(2+\epsilon D_x)D_x\left[(2-\epsilon D_x)(u-\epsilon^2u_{xx})\right]^{-\frac{2}{3}}.
 \]
 The Degasperis--Procesi equation transforms into (\ref{C4}) under the change of variables
 \[
 u\to(2-\epsilon D_x)u.
 \]

 The Lax  representation for the equation (\ref{C4}) is
 \begin{eqnarray*}
 &&\psi_x-\psi_{xxx}-\lambda\left(2m-\epsilon m_x\right)\psi=0,\quad m=u-\epsilon^2u_{xx}\\
 &&\psi_t=\frac{2}{\lambda}\psi_{xx}+2(2u-\epsilon u_x)\psi_x-2(2u_x-\epsilon u_{xx}+\frac{2}{3\lambda})\psi.
 \end{eqnarray*}

 Note that the other transformation $u\to(2+\epsilon D_x)u$ of Degasperis-Procesi gives the equation
 $(1-\epsilon^2D_x^2)u_t=D_x(2-\epsilon D_x)\left[(2+\epsilon D_x)u\right]^2$, which transforms into (\ref{C4}) under the change $x\to -x,\,\,t\to -t$.

\underline{Equation (\ref{F2}).} Equation (\ref{F2}) possesses a
hierarchy of local higher symmetries  and the first non-trivial one
is
\[
u_{\tau}=D_x\left[(1-\epsilon D_x)u\right]^{-1}.
\]
The last equation is linearisable by the transformation
\[
x=-\epsilon\log(v_y(y,t)),\quad
u=\frac{1}{\sqrt{\epsilon}\log(v(y,t))_y},\quad\Longrightarrow
v_t=v_{yy}.
\]

\underline{Equation (\ref{F4}).} The higher symmetries of this
equation are quasi-local and the first non-trivial one is
\[
(1+\epsilon D_x)u_{\tau}=D_x\left[(1-\epsilon^2D_x^2)u\right]^{-1}.
\]
However, the equation (\ref{F4}) can be rewritten as
\[
m_t=D_x(2-\epsilon D_x)\left[(1+\epsilon D_x)u\right]^2,\quad
m=u-\epsilon^2u_{xx}
\]
and the latter equation possesses an infinite hierarchy of local
higher symmetries in dynamical variable $m$. One can easily check
that the first such a symmetry is
\[
m_{\tau}=D_x(1-\epsilon D_x)m^{-1}.
\]
The last equation is linearisable by the transformation:
\[
x=-\epsilon\log(v(y,t)),\quad
m=-\frac{1}{\sqrt{\epsilon}\log(v(y,t))_y},\quad\Longrightarrow
v_t=v_{yy}.
\]

Equations (\ref{F2}) and (\ref{F4}) are related by the
transformation $u\to (1+\epsilon D_x)u$. It is clear that this
transformation does not preserve the locality of higher symmetries
of equation (\ref{F2}).

\underline{Equation (\ref{F1}).} The first non-trivial higher symmetry of this equation is quasi-local
\[
(1+\epsilon D_x)u_{\tau}=D_x\left[(2-\epsilon D_x)(u-\epsilon^2u_{xx})\right]^{-2}.
\]
However equation (\ref{F1}) can be written as
\[
m_t=D_x\left[(2-\epsilon D_x)(1+\epsilon D_x)u\right]^2,\quad m=u-\epsilon^2u_{xx}
\]
and the latter equation possesses an infinite hierarchy of local higher symmetries and the first one reads as $m_{\tau}=D_x(1-\epsilon D_x)\left[(2-\epsilon D_x)m\right]^{-2}$. The Lax representation for  equation (\ref{F1}) is not known yet.

\underline{Equation (\ref{F3}).} The first non-trivial higher symmetry of equation (\ref{F3}) is
\[
u_{\tau}=D_x\left[(2-\epsilon D_x)(1-\epsilon D_x)u\right]^{-2}.
\]
This equation possesses an infinite hierarchy of local higher symmetries. Note that equation (\ref{F1}) can be obtained from (\ref{F3}) by the transformation $u\to (1+\epsilon D_x)u$. The Lax representation for this equation is not known yet.

\underline{Equation (\ref{B3})} is a local  second order linearisable evolutionary equation \cite{MSS}, while \underline{equation (\ref{B1})} transforms into (\ref{B3}) as $u\to (1+\epsilon D_x)u$.

\underline{Equations  (\ref{pCH})-(\ref{pF3})} can be obtained from equations  (\ref{CH})-(\ref{F3}) via the
potentiation transformation $u\to u_x$.

\subsection{Equations with cubic nonlinearity}

Now we consider equations with cubic non-linearity:

\begin{The}
Suppose that  at least one of the following equations is not satisfied:
\[
c_2=0,\quad c_7=0,\quad c_{11}=0,\quad c_1+c_4=0.
\]
Then if  equation (\ref{e3}) possesses an infinite hierarchy of quasi-local
higher symmetries then up to re-scaling $x\to\alpha x, \,t\to\beta
t,u\to\gamma u$, $\alpha,\beta,\gamma=const$ it is one of the list:
\begin{eqnarray}
\label{A}
(1-\epsilon^2D_x^2)u_t&=&\epsilon^2u^2u_{xxx}+3\epsilon^2uu_xu_{xx}-4u^2u_x,\\ \nonumber \\
\label{B}
(1-\epsilon^2D_x^2)u_t&=&D_x\left(\epsilon^2u^2u_{xx}-\epsilon^4u_x^2u_{xx}+\epsilon^2uu_x^2-u^3\right),\\
\nonumber \\
\label{C}
(1-\epsilon^2D_x^2)u_t&=&\epsilon^4u_x^2u_{xxx}+\epsilon^4u_xu_{xx}^2+2\epsilon^3uu_xu_{xxx}+\epsilon^3uu_{xx}^2+\epsilon^3u_x^2u_{xx}+
\epsilon^2u^2u_{xxx}-\epsilon^2u_x^3\\ \nonumber &-&\epsilon
u^2u_{xx} -3\epsilon u
u_x^2-2u^2u_x,\\ \nonumber \\
\label{D} (1-\epsilon^2D_x^2)u_t&=&(1+\epsilon D_x)\left(\epsilon u^2u_{xx}+\epsilon uu_x^2-2u^2u_x\right),
\\ \nonumber \\
\label{E} (1-\epsilon^2D_x^2)u_t&=&(1+\epsilon D_x)\left(2\epsilon^3u_x^2u_{xx}-\epsilon^2uu_xu_{xx}-\epsilon^2u_x^3-\epsilon u^2u_{xx}-\epsilon uu_x^2+2u^2u_x\right),\\ \nonumber \\
\label{F} (1-\epsilon^2D_x^2)u_t&=&(1-\epsilon^2 D_x^2)\left(\epsilon^2u^2u_{xxx}-\epsilon^2uu_xu_{xx}+
\frac{4}{9}\epsilon^2u_x^3+cu^2u_x\right),\quad c\in\bbbc,
\end{eqnarray}
\begin{eqnarray}
\label{G}
(1-\epsilon^2D_x^2)u_t&=&(1-\epsilon^2D_x^2)\bigg(\epsilon^2u^2u_{xxx}+\epsilon^2uu_xu_{xx}-\frac{2}{9}
\epsilon^2u_x^3+c u^2u_x\bigg),\, c\in\bbbc,\\ \nonumber \\
\label{H} (1-\epsilon^2D_x^2)u_t&=&(1-\epsilon^2D_x^2)\left(\epsilon^2u^2u_{xxx}+\epsilon^2uu_xu_{xx}-\frac{2}{9}\epsilon^2u_x^3+
3c\epsilon u^2u_{xx}\right.\\ \nonumber &&\qquad\qquad\qquad\qquad\qquad\qquad\qquad\qquad\qquad\left.+c\epsilon u u_x^2+2c^2u^2u_x\right),\quad c\in\bbbc,\\ \nonumber \\
\label{I} (1-\epsilon^2D_x^2)u_t&=&(1-\epsilon^2D_x^2)\left(\epsilon^2u^2u_{xxx}+\frac{1}{9}\epsilon^2u_x^3+3c\epsilon u^2u_{xx}+c\epsilon u u_x^2+2c^2u^2u_x\right),\quad c\in\bbbc,\\ \nonumber \\
\label{J} (1-\epsilon^2D_x^2)u_t&=&(1-\epsilon^2D_x^2)\left(\epsilon u^2u_{xx}+cu^2u_x\right),\quad c\in\bbbc.
\end{eqnarray}
\end{The}
The proof requires to check the quasi-locality conditions of the first three terms of the formal recursion operator.

\underline{Equation (\ref{A}).} The first local higher symmetry of
this equation is
\[
u_{\tau}=m^{-\frac{13}{3}}\epsilon^2\left(mm_{xxx}-5m_xm_{xx}\right)+\frac{40}{9}\epsilon^2m^{-\frac{16}{3}}m_x^3-4m^{-\frac{10}{3}}m_x,\quad
m=u-\epsilon^2u_{xx}.
\]
Equation (\ref{A}) can be rewritten as
\[
m_t=-(u^2m_x+3muu_x),\quad m=u-\epsilon^2u_{xx}.
\]
The Lax representation for the equation (\ref{A}) is
\begin{eqnarray*}
\epsilon^3\psi_{xxx}&=&\epsilon\psi_x+\lambda
m^2\psi+2\epsilon^3\frac{m_x}{m}\psi_{xx}+\frac{mm_{xx}-2m_x^2}{m^2}\psi_x,\\
\psi_t&=&\frac{\epsilon}{\lambda}\frac{u}{m}\psi_{xx}-\frac{\epsilon}{\lambda}\frac{mu_x+um_x}{m^2}\psi_x-u^2\psi_x.
\end{eqnarray*}
Equation (\ref{A}) has been recently studied in detail in \cite{HW},
where the Lax representation in a different form was constructed.
The authors of \cite{HW}, \cite{HLS}  also obtained the
bi-Hamiltonian structure and constructed the peakon solutions for
equation (\ref{A}), for which the positions and amplitudes of the
peaks satisfy a finite-dimensional integrable Hamiltonian system.

\underline{Equation (\ref{B}).} The first local higher symmetry of
 equation (\ref{B}) is
\[
u_{\tau}=m^{-3}m_x,\quad
m=u-\epsilon^2u_{xx}.
\]
The equation (\ref{B}) can be rewritten as
\[
m_t=(\epsilon^2u_x^2-u^2)m_x-2m^2u_x.
\]
This equation was recently derived from shallow water theory in \cite{Qiao},
where the Lax reperesentation, and bi-Hamiltonian structure were presented and different types of solutions were constructed; however, an equivalent form of this equation was given by Fokas in \cite{F}.

\underline{Equation (\ref{C}).}  The higher symmetries of this equation are quasi-local and the first one reads as
\[
(1+\epsilon D_x)u_{\tau}=m^{-7}\left(\epsilon mm_{xx}-3\epsilon m_x^2-2mm_x\right),\quad m=u-\epsilon^2u_{xx}.
\]
Equation (\ref{C}) can be rewritten as
\[
 m_t=-\epsilon^2u_x^2m_x-2muu_x+m^2u_x-2\epsilon u u_x m_x+\frac{1}{\epsilon}mu(m-u)-\epsilon mu_x^2-u^2m_x
 \]
and the latter equation possesses an infinite hierarchy of local higher symmetries in $m$. The first such symmetry is
\[
m_{\tau}=(1-\epsilon D_x)m^{-7}\left(\epsilon mm_{xx}-3\epsilon m_x^2-2mm_x\right),\quad m=u-\epsilon^2u_{xx}.
\]

\underline{Equation (\ref{D}).} Equation (\ref{D}) possesses an
infinite hierarchy of local higher symmetries and the first
non-trivial one is
\[
u_{\tau}=v^{-7}\left(\epsilon vv_{xx}-3\epsilon
v_x^2-2vv_x\right),\quad v=u-\epsilon u_x.
\]

\underline{Equation (\ref{E})}. The first local higher symmetry of this equation is
\[
u_{\tau}=v^{-2}(v+\epsilon v_x)^{-1}-v^{-3},\quad v=u-\epsilon u_x.
\]
The latter equation is linearisable as it a second order integrable evolution equation (cf. equations  (\ref{F2}), (\ref{F4}). )

\underline{Equations (\ref{G})-(\ref{J})} correspond to local evolutionary equations of orders 3 and 2.

\section{Conclusions}

In this paper we have considered polynomial homogeneous
generalisations  of the Camassa--Holm type equation with quadratic
and cubic nonlinearity. We have classified all  equations of the
form (\ref{e1}), (\ref{e2}), (\ref{e3}), which possess infinite
hierarchies of (quasi)-local higher symmetries. We have shown that
the obtained equations can be treated as non-local symmetries of
local scalar evolution quasi-linear integrable equations of orders
2, 3 and 5.

Some of the obtained equations seem to be new and are likely to
provide more examples of solution phenomena (peakons, compactons,
other weak/non-classical solutions) that do not appear in local
evolution equations \cite{LOR}. The study of multi-phase solutions
of these equations remains out of the scope of this paper.

The author is very grateful to A.N.W. Hone, A.V. Mikhailov, J.P.
Wang, E.V. Ferapontov, R. Camassa for valuable discussions. This
work was supported by EPSRC Postdoctoral Fellowship C/527747/1.

\end{document}